\documentstyle[onecolumn,epsfig]{mn}
\title{Phase Information and the Evolution of Cosmological Density Perturbations}

\author[Chiang \& Coles] {Lung-Yih Chiang$^{1}$ and Peter
Coles$^{1,2}$\\ $^1$ Astronomy Unit, School of Mathematical
Sciences, Queen Mary \& Westfield College, London E1 4NS, UK.
\\ $^2$ School of Physics \& Astronomy, University of Nottingham,
University Park, Nottingham NG7 2RD, UK.}

\begin{document}

\maketitle

\begin{abstract}
The Fourier transform of cosmological density perturbations can be
represented in terms of amplitudes and phases for each Fourier
mode. We investigate the phase evolution of these modes using a
mixture of analytical and numerical techniques. Using a toy model
of one-dimensional perturbations evolving under the Zel'dovich
approximation as an initial motivation, we develop a statistic
that quantifies the information content of the distribution of
phases. Using numerical simulations beginning with more realistic
Gaussian random-phase initial conditions, we show that the
information content of the phases grows from zero in the initial
conditions, first slowly and then rapidly when structures become
non-linear. This growth of phase information can be expressed in
terms of an effective entropy: Gaussian initial conditions are a
maximum entropy realisation of the initial power-spectrum,
gravitational evolution {\em decreases} the phase entropy. We show
that our definition of phase entropy results in  a statistic that
explicitly quantifies the information stored in the phases of
density perturbations (rather than their amplitudes) and that this
statistic displays interesting scaling behaviour for self-similar
initial conditions.
\end{abstract}


\section{Introduction}

Observations of large-scale structure in the spatial distribution
of galaxies pose some of the most interesting challenges facing
modern cosmological theory. Prominent among the issues raised by
the confrontation is the question of how to best to use the
information contained in maps of galaxy positions to constrain
theoretical models.
\begin{figure}
\caption{A simple
demonstration of the importance of phase information for pattern
morphology. Panel (a) shows an example realisation of an N-body
experiment with initial random phases. Panel (b) is obtained by
taking the  Fourier transform of panel (a), setting all the
amplitudes to a constant ($|\tilde{\delta}_k|=1$) then taking the
inverse Fourier transform; it therefore retains only the phase
information from (a). In panel (c), each mode keeps the same
amplitude so the power-spectrum is unchanged but we redistribute
the phases randomly among the modes. It is easy to see the
striking resemblance between (a)and (b), the phase-only
reconstruction, but (c) which has the same power-spectrum but
random phases, is featureless. }
\end{figure}

Traditional methods for the analysis of such maps involves a
Fourier decomposition, treating the cosmological density contrast
as a superposition of plane waves with (complex) amplitudes
$\tilde{\delta}_k$:
\begin{equation}
\delta ({\bf x}) = \frac{\rho({\bf x})-\rho_0}{\rho_0}=\sum
\tilde{\delta}_k \exp(i{\bf k}\cdot {\bf x}),
\end{equation}
where $\rho_0$ is the mean matter density. We discuss this kind of
spectral representation in more detail in Section 4. The
statistical analysis of galaxy clustering generally proceeds via
the properties of the amplitudes $\tilde{\delta}_k$. A
particularly useful statistical quantity in this respect is the
power spectrum, essentially proportional to
$|\tilde{\delta}_k|^2$, which is a second-order statistic in
Fourier space. Higher order quantities based on $\tilde{\delta}_k$
can also be found, such as the bispectrum (Peebles 1980;
Matarrese, Verde \& Heavens 1997) or through correlations of
$\tilde{\delta}_k^2$ (e.g. Stirling \& Peacock 1996). However, an
alternative approach exists that has not been explored greatly in
the cosmological literature, and that is to look explicitly not at
the amplitudes (the moduli of $\tilde{\delta}_k$) but their {\em
phases} $\phi_k$ where
\begin{equation}
\tilde{\delta}_k=|\tilde{\delta_k}|\exp(i\phi_k).
\label{eq:fourierex}
\end{equation}

The importance of phase information can be understood by
considering a simple example. Gaussian white noise consists of a
superposition of Fourier modes with random phases and whose
amplitudes are drawn from a distribution which is independent of
wavenumber, i.e. a flat spectrum. Now imagine a density field
represented as a single Dirac $\delta$-function in a periodic
volume. The Fourier transform of this density field is constant in
amplitude, so that its power spectrum corresponds exactly to the
white noise spectrum. What differs drastically in these two cases
is the difference in phases. In the former, the phases are random,
in the latter, the phase of each mode is a definite relationship
to the other modes resulting in a real-space distribution that is
strongly localised. This second distribution clearly represents a
more {\em ordered} state than the first one, but their power
spectra are identical. A more relevant example is shown in Figure
1. The left panel shows a grey-scale representation of the density
field obtained from a two-dimensional N-body simulation. In the
centre panel, the amplitudes of the Fourier components are set to
a constant value, but the phases are retained. The morphology of
the density field is very similar to the original. The result of
randomly reshuffling the phases but retaining the amplitudes is
shown on the right; information about the shape of the original
pattern is destroyed. One can conclude from these examples that
key information about the shape of localised features in a spatial
pattern resides in the distribution of phases, and that
information should in principle be quantifiable. Further
interesting general examples of the importance of phase
information for the analysis of images (in the widest sense of the
word) can be found in Oppenheim \& Lim (1981). The question is,
how to quantify the information content of the distribution of
phases?

The density distributions of interest in cosmology are, of course,
more complicated than these toy examples. Moreover, since
gravitational clustering is generally thought to be driven by the
process of gravitational instability we need to understand not
only the phase information in ``snapshots'' of the density field,
but the way in which phase information evolves with time.

The standard model for the formation of galaxies and large-scale
structure involves the idea that these structures grew by the
action of gravitational instability from small initial density
perturbations present in the early Universe. In most popular
variants of this model, particularly those involving cosmic
inflation, the initial perturbations are Gaussian, meaning amongst
other things that their probabilistic properties are completely
specified by knowledge of second-order statistical quantity: the
autocovariance function, or its Fourier transform the power
spectrum discussed above. Gaussian random fields are useful not
only because of this very economical prescription of their
statistical properties, but also because many properties of
Gaussian random density fields can be calculated analytically
(e.g. Bardeen et al. 1986). One particularly interesting property
of Gaussian random fields is that they possess Fourier modes whose
real and imaginary parts are independently distributed or, in
other words, that they have phases which are independently
distributed and uniformly random on the interval $[0,2\pi]$. Terms
in the evolution equations for the Fourier modes that represent
coupling between different modes are of second (or higher) order
in $\delta$ and these are neglected when first-order perturbation
theory  is considered. During the linear regime, therefore,
Fourier modes evolve independently and the phases remain
independent and uniformly random (Peebles 1980; Sahni \& Coles
1995). In the later stages of evolution, however, modes begin to
couple together. It is in this regime that information flows into
the set of Fourier phases.

There are therefore two reasons for studying phase correlations.
One is to understand exactly how the morphology of non-linear
clustering pattern is driven by the growth of phase information.
The other is to explore ways of testing the initial hypothesis of
Gaussian initial conditions, using measurements of phase
association to constrain the level of possible non-Gaussianity.

Some papers on this subject refer to the appearance of phase
information in terms of {\em phase correlations}. In a strict
sense of the word correlation, which means association expressed
through second-order quantities such as covariances, this is
incorrect. As we show in Section 4, any statistically homogeneous
field cannot display phase correlations. However, it is certainly
true that phases are no longer independently distributed in this
regime.

Despite the clear importance of phases for the morphology of
galaxy clustering, the relevant literature is relatively sparse,
with most attention being focused on the evolution of individual
phases away from their initial values (Ryden \& Gramman 1991; Soda
\& Suto 1992; Jain \& Bertschinger 1998). One notable exception is
the work by Scherrer, Melott \& Shandarin (1991) who suggested and
explored a method of quantifying phase association that could be
applied to real data. The sparsity of the existing literature on
phases is at least partly due to the difficulty in developing
statistical methods for quantifying phase information, a
difficulty we attempt to address in this work.

In this paper we discuss the phase evolution and the
quantification of the information they represent. After presenting
some relevant background in Section 2, we begin by exploring the
evolution of phases using simplified 1D analytical models in
Section 3. This study suggests a simple way of encoding phase
information which we present in Section 5. With the
aid of dynamical numerical simulations in Section 6, we present how the encoded information from
phases evolves with different initial conditions and displays interesting behaviour for
self-similar evolutions. In Section 7,
we discuss the results and offer suggestions on how phase
information might be extracted from real surveys.

\section{Theoretical Background}
\subsection{Linear Perturbation Theory}
For the purposes of this study we consider a spatially flat
matter-dominated universe. If the mean free path of the particles
is small, it is appropriate to treat the large-scale distribution
of  matter as a self-gravitating Newtonian fluid with zero
pressure. In this case the evolution of the gravitating system can
be described in a comoving frame by the following three equations:
the  {\em continuity equation},
\begin{equation}
\frac{\partial \delta }{\partial t}+\frac{1}{a}\nabla_{\bf
x}[(1+\delta){\bf v}=0;   \label{eq:continuity}
\end{equation}
Euler equation,
\begin{equation}
\frac{\partial(a{\bf v})}{\partial t}+({\bf v}\cdot \nabla_{\bf
x}){\bf v}=-\frac{1}{\rho}\nabla_{\bf x}p-\nabla_{\bf x}\phi;
\label{eq:euler}
\end{equation}
and Poisson equation,
\begin{equation}
\nabla^{2}_{\bf x}\phi = 4 \pi G a^{2} \rho_{0} \delta.
\label{eq:poisson}
\end{equation}
In these equations $\delta({\bf x})$ is the density contrast
defined by eq. (1); for a spatially flat and matter-dominated
universe the mean density $\rho_{0}=1/6\pi G t^{2}$. In the
equations (\ref{eq:continuity}-\ref{eq:poisson}), $\nabla_{\bf x}$
denotes a derivative with respect to  the comoving coordinates
${\bf x}$, where ${\bf x}={\bf r}/a(t)$, and ${\bf r}$ are proper
coordinates. The velocity field is ${\bf v}({\bf x},t)=a \dot{\bf
x}$ and $\phi({\bf x},t)$ is the peculiar gravitational potential.
The regime in which  $ \delta\ll 1$, permits one to linearise
(\ref{eq:continuity}) and (\ref{eq:euler}) by keeping only terms
of first order in the perturbed quantities. We thus obtain
\begin{equation}
\frac{\partial^{2}\delta}{\partial t^{2}}+
2\left(\frac{\dot{a}}{a}\right)\frac{\partial \delta}{\partial
t}-4 \pi G \rho_{0} \delta=0.
\end{equation}
Solving this equation with the initial condition $\delta({\bf
x},t_{i})=\delta_{0}({\bf x})$  and $\dot{\delta}({\bf
x},t_{i})=0$ at $t_{i}$, when the perturbations start growing, we
get
\begin{equation}
\delta({\bf x},t)=\delta_{0}b_{\pm}(t), \label{eq:linearsol}
\end{equation}
with the growing mode, $b_{+}\propto t^{2/3}$,  and the decaying
mode, $b_{-}\propto t^{-1}$. This solution describes the growth of
perturbations at early times. In this regime the evolution of each
individual Fourier mode, $\tilde{\delta_{k}}$, of $\delta$ is
decoupled from the other modes and its rate of growth is
independent of $k$.

The linear solution breaks down when $\delta$ is compatible with
unity or beyond, because terms of higher order than the linear
terms become important. To probe the onset of non-linearity we
therefore need to use more sophisticated approximation methods
than first-order perturbation theory (see, e.g. Sahni \& Coles
1995, for a thorough review).

\subsection{The Zel'dovich Approximation}
Zel'dovich (1970) proposed a particularly ingenious approximate
scheme which can be used to extrapolate the evolution of density
fluctuations well into the non-linear regime. In the eponymous
Zel'dovich approximation (ZA), a particle initially placed at
Lagrangian coordinate ${\bf q}$  finds itself after a time $t$ at
an Eulerian coordinate  is ${\bf x}$. The displacement it has
experienced simply depends on the velocity the particle had when
it was at its original position, through $b(t){\bf u}({\bf q})$,
so that
\begin{equation}
{\bf r}({\bf q},t)=a(t){\bf x}({\bf q},t)=a(t)[{\bf q}+b(t){\bf u}({\bf q})],
\end{equation}
where  $a(t)$ is the expansion factor and $b(t)$ is the growing
mode $b_{+}$ in (\ref{eq:linearsol}). According to this
prescription, the motion of the medium is simply Newtonian
inertial motion: each particle moves with its original velocity
along a ballistic trajectory. [Note that the peculiar velocity
according to the  Zel'dovich approximation is $a\dot{\bf
x}=a(t)\dot{b}(t){\bf u}({\bf q})$.] For an irrotational flow,
${\bf u}({\bf q})$ can be expressed as a gradient of some velocity
potential $-\nabla \Phi_{0}({\bf q})$.

The evolutionary picture that the  Zel'dovich approximation
suggests is that particles set out on  inertial trajectories with
their initial velocities determined by the initial velocity
potential $\Phi_{0}({\bf q})$  and follow straight lines with
constant velocities. When two such trajectories meet, a
singularity forms and both their positions, being distinct in
Lagrangian space, are mapped onto the same Eulerian coordinate.
Such a singularity is called a {\em caustic} and its formation is
called {\em shell-crossing}. Because the Zel'dovich approximation
does not deal properly (or, indeed, at all) with the self-gravity
of regions formed when such caustics occur, the two particles
entering the singularity do not coalesce in a bound structure, but
their paths cross each other and form multi-stream flows. The
approximation breaks down at this point, and the mapping from
Lagrangian to Eulerian space can no longer be inverted.
Nevertheless, before shell crossing, the Zel'dovich approximation
provides an excellent approximation to the real evolution of the
density contrast (e.g. Coles, Melott \& Shandarin 1993).

Before shell-crossing the evolution of $\delta$ can be determined
which can be obtained by requiring mass conservation between
Lagrangian and Eulerian space
\begin{equation}
\rho_{0} \, d^{3}q=\rho({\bf x},t) \, d^{3}x.
\end{equation}
If the flow is assumed irrotational, the matrix $\partial
x_{i}/\partial q_{j}$ is a symmetric matrix and can be
diagonalised. Therefore,
\begin{equation}
\rho({\bf x},t)=\frac{\rho_{0}}{(1-b(t)\lambda_{1})(1-b(t)\lambda_{2})
(1-b(t)\lambda_{3})},
\end{equation}
if $-\lambda_{1}$, $-\lambda_{2}$, $-\lambda_{3}$ are the
eigenvalues of $ \partial u_{i}(q)/ \partial q_{j}$. Thus, the
condition that a caustic forms is the same condition as that one
eigenvalues of the matrix $\partial x_{i}/\partial q_{j}$ should
be equal to $1/b(t)$.

The Zel'dovich approximation breaks down entirely in the highly
non-linear regime, but this is not too great a problem for the
problem at hand. As other studies of phase evolution have shown
(Ryden \& Gramman 1991; Scherrer, Melott \& Shandarin 1991; Soda
\& Suto 1992; Jain \& Bertschinger 1998), the phases of Fourier
modes $\phi_k$ begin to move very rapidly away from their original
values $\phi_i$ when density fluctuations become very large. This
poses a problem for the interpretation of phase information when
each one wraps around many multiples of $2\pi$. In fact, what
happens in the non-linear regime is that phases move so quickly
for short-wavelength modes that their displacements become very
much larger than $2\pi$. The `observed' phases, measured modulo
$2\pi$, then appear random because of the heavy phase wrapping in
the strongly non-linear regime. This is is a difficult problem to
deal with, so we will simply skirt around it and consider only
what happens on scales just entering the non-linear regime.

\subsection{The Zel'dovich Approximation in One Dimension}
Although the Zel'dovich approximation (ZA) is a very useful tool
for following the fully three-dimensional evolution of density
perturbations, it is particularly noteworthy that the solution it
provides in one dimension is exact, at least until the first
shell-crossing. In one dimension, including the case of a single
plane wave in 3D, the value of $\delta$ calculated from the
Poisson equation is the same as that found using the ZA. The ZA is
therefore a particularly useful tool to explore the evolution of
the density contrast in 1D; every particle carries information
about its position and velocity during the evolution. In one
dimension, the ZA is simply
\begin{equation}
x(q,t)=q-b(t)\frac{d \Phi_{0}(q)}{dq},
\end{equation}
and the density contrast is
\begin{equation}
\delta=\left(\frac{\partial x}{\partial q}\right)^{-1}-1=\left(1-b(t)\frac{d^{2}
\Phi_{0}(q)}{dq^{2}}\right)^{-1}-1, \label{eq:delta}
\end{equation}
until the first singularity develops, at which point
\begin{equation}
\frac{dx}{dq}=0, \end{equation} or \begin{equation}
b(t)=\left(\frac{d^{2}\Phi_{0}(q)}{dq^{2}}\right)^{-1}_{\rm min}. \label{eq:shellcrossing}
\end{equation}
The  velocity potential field is closely related to gravitational
potential field, but it is important to realise that this is only
relevant to the initial velocities of particles and not to their
final positions; density peaks take place, not at the minima of
potential wells, at the positions where the change of curvature in
the potential field is zero, i.e.\ $d^{3}\Phi_{0}/dq^{3}=0$ with
$d^{4}\Phi_{0}/dq^{4}$ at their minima.

The fact that the ZA is exact in one dimension until
shell-crossing allows us to study this case with particular ease
(cf. Soda \& Suto 1992). What we shall do in the next section is
to study the evolution of phases in 1D in order to gain at least a
qualitative understanding of how non-linear evolution leads to an
increase in the information content of the set of Fourier phases.
These calculations represent, for the time being, toy models, but
as we shall see they will lead to the suggestion of a very useful
way of quantifying phase evolution in the more general case.

\section{One-dimensional Analytic Calculations}

The Fourier transform  of the density contrast $\delta(x)$ in one
dimension can be expressed as
\begin{equation}
\tilde{\delta_{k}}=\frac{1}{2\pi}\int^{\pi}_{-\pi}\delta(x)\,\exp(ikx)\,dx=
\Delta_{k}e^{i\phi_{k}}, \label{eq:fourier}
\end{equation}
where $\phi_{k}$ is the phase of the $k$-th mode, and
$\Delta_{k}=|\tilde{\delta}_k|$, which is real, is its amplitude;
c.f. equation (\ref{eq:fourierex}). Here it is assumed that the universe is
periodic in one dimension with a period of $2\pi$, i.e.\,
$x(q+2j\pi)=x(q)$. Typically, we can represent the velocity
potential as the  sum of sinusoidal functions, i.e.
\begin{equation}
\Phi_{0}(q)=\sum_{k}A_{k}\cos(kq+\alpha_{k});
\end{equation}
the assumption of periodicity requires that $k$ here must be an
integer.

Getting an analytic solution for the Fourier modes of the density
distribution evolved according to the Zel'dovich approximation is
not  as straightforward as one might think because the Fourier
transform (\ref{eq:fourier}) is made from real-space Eulerian
coordinate, which in turn represents a mapping from the
Lagrangian coordinate. From (\ref{eq:fourier})
\begin{equation}
\delta_{k}=\frac{1}{2\pi}\int^{\pi}_{-\pi}
\left[\left(\frac{dx}{dq}\right)^{-1}-1\right]e^{ikx}
dx=\frac{1}{2\pi}\int^{\pi}_{-\pi}e^{ikx}
dq=\frac{1}{2\pi}\int^{\pi}_{-\pi}\exp\left[ik\left(q-b(t)
\frac{d\Phi_{0}(q)}{dq}\right)\right]dq.
\end{equation}

Obviously, the case of an arbitrary $\delta(x)$ comprising a
number of different Fourier modes is going to be difficult to
analyse, so we will begin by looking at simpler cases.

\subsection{Single Mode}
The simplest case we can consider is one in which the velocity
potential is described by a velocity potential comprising a single
sinusoidal function:
\begin{equation}
 \Phi_{0}(q)=-\cos(q+\alpha).
\end{equation}
An analytical solution is available in this case
\begin{equation}
\tilde{\delta_{k}}=\frac{1}{2\pi}\int^{\pi}_{-\pi}dq\,\exp\{ik[q-b(t)\,\sin(q+\alpha)]\}
=\exp(-ik\alpha){\it J}_{k}[kb(t)],
\end{equation}
where ${\it J}_{k}[kb(t)]$  is the {\it Bessel} function of the $k$-th
order and the phase $\phi_{k}=-k\alpha$ (Shandarin \& Zel'dovich
1989). When $b(t)\ll 1$, $\Delta_{k}\propto [kb(t)]^{k}$,
indicating the amplitudes of low-frequency modes  grow faster than
those of high-frequency ones. As $b(t)$ is approaching unity, the time
when the first caustic forms, the amplitude for $k=1$ slows down,
compared with the extrapolation of the linear theory (in which the
amplitudes grow with the same rate). The phases, on the other
hand, remain the same until shell-crossing. This idealised case
demonstrates that the collapse of a single plane wave involves the
generation of higher-frequency modes with phases determined by the
phase of the original mode.

\subsection{Double Mode}
The next case we consider is where the initial velocity potential
is simply the superposition of two sinusoidal forms:
\begin{equation}
 \Phi_{0}(q)=-\cos(k_{1}q)-\cos(k_{2}+\alpha).
\end{equation}
Even in this very simple case, no analytical solution can be
obtained. We begin with a particular choice of frequencies for the
two starting modes, one being the fundamental mode $k_{1}=1$ and
the other being double this frequency. In Figure~2, we show five
different values of $\alpha$ for $k_{1}=1$ and $k_{2}=2$:
$\alpha=0.3$, 1.1, 2.0, 2.8, and 3.1. The first row shows the
corresponding velocity potential. There are two wells in the
potential field; consequently two density peaks form during the
later stages of evolution, shown in the second row. The density
contrast $\delta$ is plotted against $q$ when $b(t)=0.2$ for all
five $\alpha$'s. The choice of this $b(t)$ value is arbitrary but
to show density contrast at very late stage from which phase
coupling in Fourier domain can be explored. According to equation
(\ref{eq:shellcrossing}), different velocity potential fields have
different times for the first shell-crossing, so $\delta$ in
Figure~2 shows different levels of density peaks. In order to
elucidate the pattern of phases generated by the evolution of this
system, we use a technique borrowed from the idea of a {\em return
map} used in chaotic dynamics (e.g. May 1976).

\begin{figure}
\epsfig{file=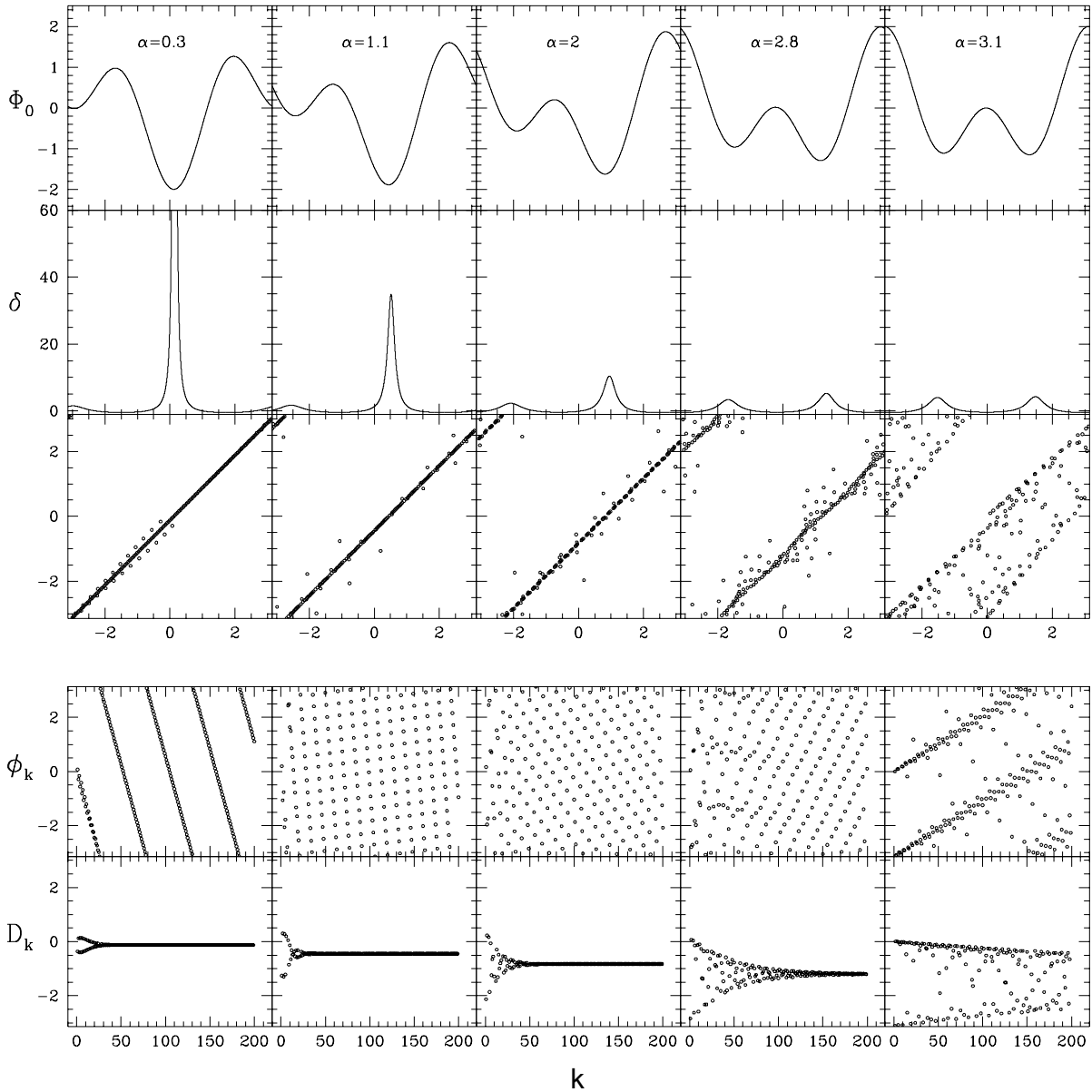,width=17cm} \caption{Double-mode
calculations with $(k_{1},k_{2})=(1,2)$. Five different values of
$\alpha$ are drawn for one-dimensional ZA solutions with an
initial velocity potential $\Phi_{0}=-\cos(q)-\cos(2q+\alpha)$.
There are two potential wells in $\Phi_{0}$, causing two density
peaks to form. The second row is the density contrast $\delta$
when $b(t)=0.2$. Both $\Phi_{0}$ and $\delta$ are periodic between
$-\pi$ and $\pi$, and are shown as functions of the Lagrangian
coordinate $q$. The third row is the corresponding return map,
$\phi_{k+1}$ versus $\phi_{k}$, both of which are measured modulo
$2\pi$. The fourth and fifth rows are the phase $\phi_k$ and
discrete phase gradient $D_k$, respectively, both being drawn
against  $k$. Histograms of the phases $\phi_k$ show a uniform
distribution, whereas in the return maps the coupling of
consecutive phases is clear. The relationship between the
formation of density contrast and the way phases couple each other
is easy to see via $D_k$ panels. A dominant density peak converges
$D_k$ on large $k$, whereas the shape of $D_k$ on small $k$'s
depends on the morphology of the structure. If two compatible
density peaks form at the same time, $D_k$ converges slowly
towards large $k$'s.}
\end{figure}

To analyse a chaotic time series $X(t)$, sampled at discrete times
$t$ (which we assume to be represented by the integers), it is
useful to plot $X_{t+1}$ against $X_{t}$ for each pair of
consecutive values $(t, t+1)$; this is the return map. If $X_t$
and $X_{t+1}$ are independent then the result should be a scatter
plot in which no pattern appears. A time series consisting of
correlated noise, would show some correlation in its return map,
whereas a chaotic series would exhibit some form of attractor. We
adapt this idea to the case in hand, by considering the sequence
of phases $\phi_{k}$ in the same way. For a $\phi_k$ sequence of a density distribution, $\phi_k$
are paired for all modes as points $(\phi_k,\phi_{k+1})$ in a return map so that one can hope
to see some pattern emerging in the relationship between two
successive phases. This idea can, in principle, be generalised to
higher-order maps showing $\phi_{k+2}$ against $\phi_{k}$, etc.

For the case of the single mode, the $\phi_{k}$ sequence decreases
monotonically by fixed quantity $\alpha$ for each step in $k$. In
the return map, the phases are mapped into points
$\{(\phi_{1},\phi_{2})=(-\alpha,-2\alpha)\),
\((\phi_{2},\phi_{3})= (-2\alpha,-3\alpha)\ldots\}$, which forms
in the return map a diagonal line, provided that $\alpha$ is not a
rational multiple or fraction of $\pi$.

\begin{figure}
\epsfig{file=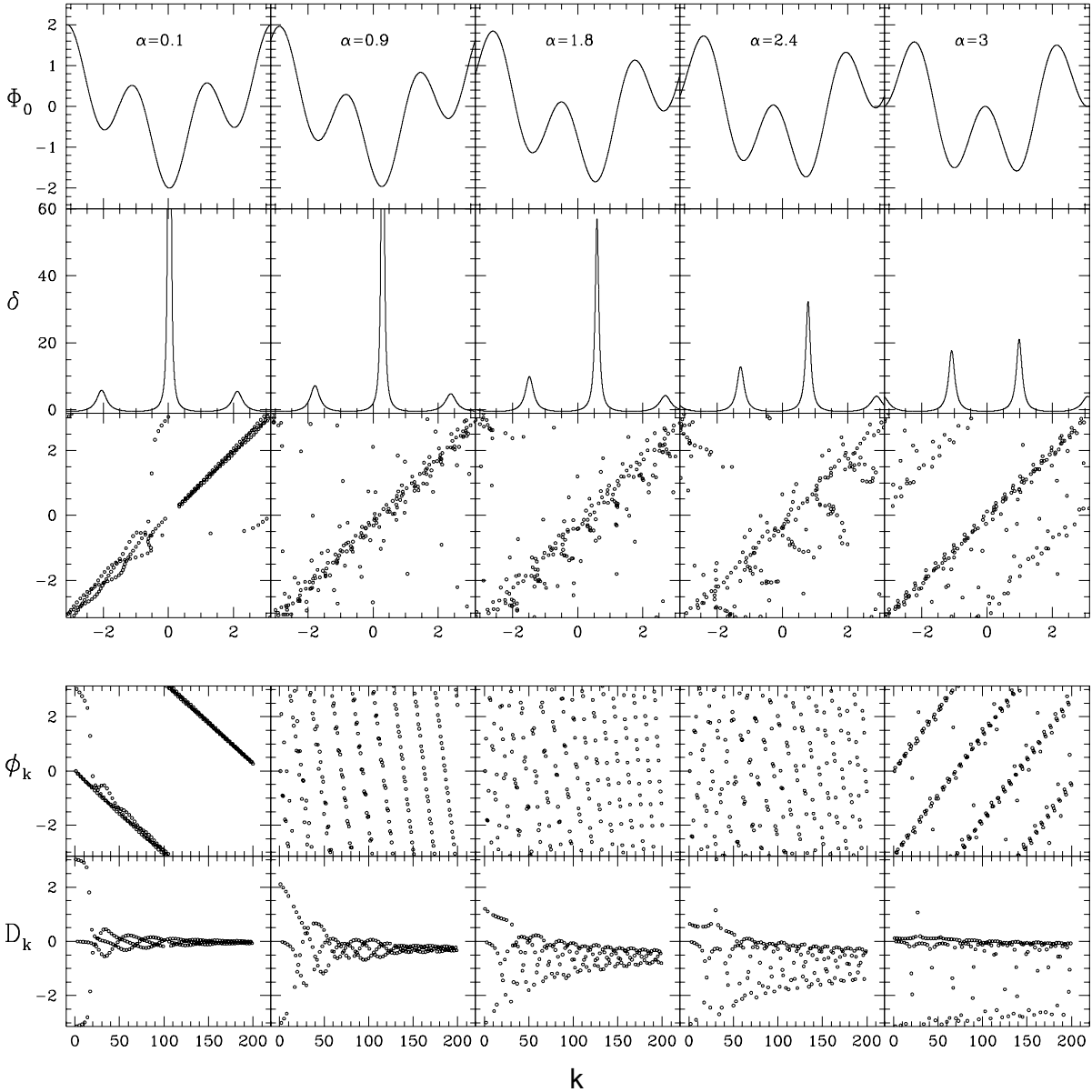,width=17cm} \caption{$(k_1,k_2)=(1,3)$ for
double mode. All the notations follow those in Figure~2. Here
three potential wells cause three density peaks. $\delta$ is drawn
when $b(t)=0.1$. The return map is more complicated in this case,
but $D_k$ plots show the same relationship between consecutive
phases as that in Figure~2. A dominant peak causes $D_{k}$ to
converge more quickly towards large $k$ than two compatible ones.}
\end{figure}

Before examining the possible patterns in the return maps, we show
in the fourth row of Figure~2 the phase $\phi_k$ as a function of
$k$. Even for the simplest case of two density peaks, the phases
display a complicated behaviour among these five different values
of  $\alpha$. In the return maps, however, it is clear to see the
patterns for the corresponding density distributions. The diagonal
lines in the first  three panels are from the mapping of
consecutive phases of high-frequency modes, which resemble the
single-mode case. The points oscillating with the diagonal lines
are from phases of long-wavelength modes. When $\alpha$ approaches $\pi$, the
phases are mapped into regions, instead of one single line.

A similar effect is noticed in Figure 3. Again, five different
potentials with different values of $\alpha$  are drawn for
$k_1=1$, and $k_2=3$. All the notations follow those in Figure~2.
Three potential wells cause three density peaks when $b(t)=0.1$.
We see again in the $\phi_k$ panels phases seem randomly
distributed as in Figure~2. In the return maps, phases are not
mapped onto diagonal lines, but much more complicated patterns. As
the patterns are the mapping from $\phi_k$ to $\phi_{k+1}$, we
devise a new quantity to look for how the consecutive phases are
coupled
 \begin{equation} D_k = \phi_{k+1}-\phi_{k}, \label{eq:defD}
\end{equation}
which means that $D_k$ is effectively a discrete phase gradient.
As will become clear in the next few sections, a histogram from $D_k$ can be used to
quantify phase information. Plotting a histogram of the $\phi_k$, however, is a useless
diagnostic of phase correlations. Such a plot would involve the
projection of the return map onto one or other of the axes. Even
if $\phi_k$ and $\phi_{k+1}$ were linearly related, the histogram
of phases would be uniform, as can be seen from Figure 2.
Instead of taking $D_k$ series into the return map, we plot the
discrete phase gradient $D_k$ against $k$ in the fifth row both in
Figure~2 and Figure~3. In Figure~2, the successive phases are
coupled in such a way that the series of  $D_k|_{k=1,3,5\ldots}$
and $D_k|_{k=2,4,6\ldots}$ twine together and oscillate embedded
in a decay, and in Figure~3, it is $D_k|_{k=1,4,7\ldots}$,
$D_k|_{k=2,5,8\ldots}$ and $D_k|_{k=3,6,9\ldots}$ that shows a
definite relationship between phases.

Since phases alone can recover the morphology of the structure,
what is shown in $D_k$ for $(k_1,k_2)=(1,2)$ case can be
demonstrated and understood qualitatively by two Gaussian waves,
or three for$(k_1,k_2)=(1,3)$, the reason being Gaussian function
is easy to manipulate between real and Fourier domain. It turns
out that the resulting  $\tilde{\delta}_k$ of mode $k$ is from
vector superposition in the complex plane, i.e.,
$\tilde{\delta}_k^{(1)}+\tilde{\delta}_k^{(2)}$, both being the
$k$-th mode of Fourier transform of two density peaks. The number
of twining series was caused by the number of density peaks. When
$\alpha=\pi$, two identical peaks located at $q_1$ and $-q_1$ form
symmetrically to the origin in real domain, so phases of the
$k$-th mode is formed by superposition in the complex plane of two
vectors with same amplitude, whose phases are $kq_1$ and $-kq_1$,
multiples of the fundamental mode and always symmetric to the
$x$-axis. This results in $\phi_k=0$ or $\pi$, so $D_k$ flip
between 0 and $\pi$. Bearing this picture in mind, we can
understand qualitatively why, when $\alpha$ is small, the dominant
sharp peak produces fast convergence of $D_k$ on high-frequency
modes, similar to the single-mode solution, whereas the twining
part of $D_k$ depends on the morphology of the distribution. The
vector in the complex plane representing the dominant sharp peak
has much larger amplitude than the other one. This fast
convergence of $D_k$ results from receding influence of vector
superposition on the resulting $\phi_k$ from the vector which
represents the less dominant density wave. Moreover, when $\alpha$
is tuned approaching $\pi$, a dominant sharp peak is replaced  by
two compatible ones, whose locations  are shifting towards
symmetry to the origin in real domain, so two compatible vectors
in the complex plane account for the slow convergence in $D_k$.
Therefore, the degree of convergence in $D_k$ depends on how
asymmetrical are the representional vectors for the sinusoids are
in the complex plane.

\begin{figure}
\epsfig{file=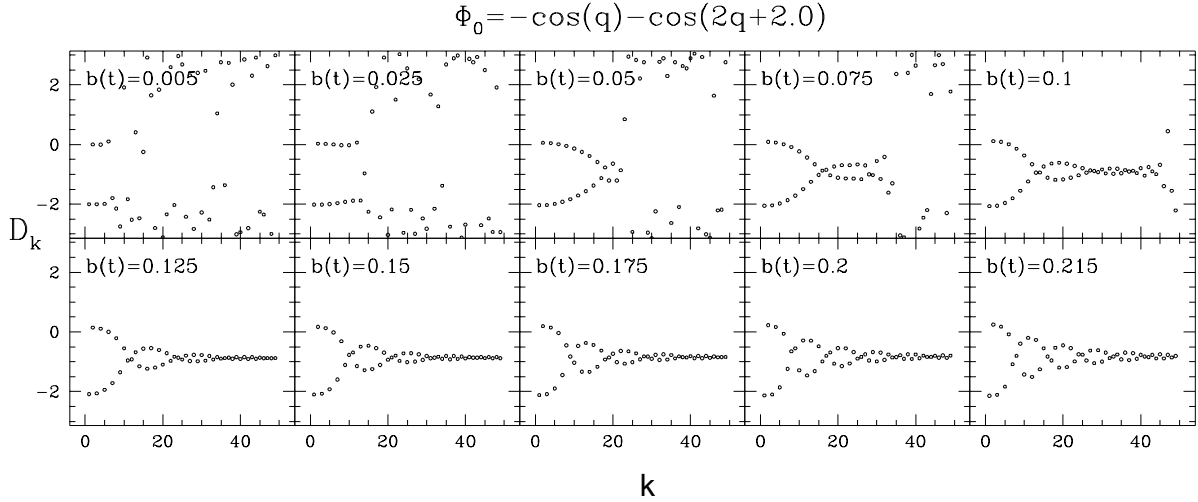,width=17cm} \caption{The evolution of
discrete phase gradient for $\Phi_{0}=-\cos(q)-\cos(2q+2.0)$. The
phase gradient $D_{k}$ is shown as a function of $k$ up to $k=50$.
In the top left panel, phases from small $k$'s start coupling at
the beginning of the evolution, which depends on the initial phase
difference between the two sinusoids,  whereas those of
high-frequency modes, whose amplitudes at this stage are still
small or zero, are rounding errors. As the density flucuation
grows, fine details appear when higher harmonics are spawned by
the collapsing waves with a definite phase relationship leading to
sharp peaks that eventually become caustics. }
\end{figure}

Figure~4 shows the evolution of the discrete phase gradient for
the case of $\Phi_0=-\cos(q)-\cos(2q+2.0)$. At the initial
condition, only two modes have well-defined amplitudes and so only
two phases really exist. However, rounding errors in the
evaluation of the required integrals result in the assignment of a
definite phase even to those modes which have formally zero
amplitude. These fictional phases are initially random when the
evolution begins. The first structure appeared in $D_k$ depends on
the phase difference between the two sinusoids, in which $D_k$
from low-frequency modes flip between 0 and $-\alpha$. Fine
details of the structure appear when higher harmonics are spawned
by the collapsing waves with a definite phase relationship leading
to sharp peaks that eventually become caustics. Those spawned
modes not only produce the twining characteristic in $D_k$, but
also interfere with the phases of low-frequency modes.

\subsection{Comments}

We have analysed these simple cases of two interacting plane waves
in an attempt to develop a quantitative understanding of the kind
of structure one can hope to see in the phases of non-linearly
evolved density fields.

The results obtained depend very sensitively on the initial phase
difference. The reason for this is that different real-space
processes become confused when viewed in the Fourier domain,
particularly in the phase representation. For one thing, each
initial sinusoid begins to collapse in the same manner as the
single-mode solution described in \S 3.1. This spawns Fourier
modes with a definite phase relationship to the initial mode. This
process generates high-frequency modes which interfere and produce
a pattern in the return map that depends quite sensitively on the
symmetry of the situation, i.e. on the initial relative phase of
the two modes. If higher-frequency modes like this had actually
been present in the initial conditions they would be affected by
the generation of higher-frequency modes by the collapse of long
wavelength perturbations. Obviously, there is no point in
continuing the discussion to the interaction between three, four
and $n$-modes.

But despite this complexity, what these examples showis that the
relatively simple quantity $D_k$, defined by eq. (21), shows an
interesting response to the non-linearity induced by mode coupling
and which quantitatively  encodes some of the structure seen in
the return maps. In Section 5 we discuss how to exploit the
information contained in this simple statistic in the analysis of
more realistic distributions obtained from an arbitrary
superposition of initial modes, rather than just two. We shall see
that the clues obtained from the illustrative examples above
suggest a new and potentially powerful method for the analysis of
phase information.

\section{Phases for Random Processes}
The previous section dealt with toy examples involving initial
perturbations that were periodic on a given interval. In
cosmology, the initial density field will not be periodic but will
be in the form of one realisation of a random process. There are
some subtleties involved in defining amplitudes and phases for
such perturbations and these are often glossed over in the
cosmological literature, although they are well known in the
literature pertaining to random processes (e.g. Priestley 1981),
so we will take the opportunity here to outline these problems and
their resolution fairly carefully. In particular, a realisation of
a random process can formally be expressed neither by a Fourier
series (which assumes periodicity) nor a Fourier integral (which
requires the function to tend to zero at $\pm \infty$). This
section is about the correct mathematical way of treating the
spectral representation of such a function, following the ideas of
Wiener (1930).

To keep things as simple as possible and consistent with the
previous section, consider a process $\delta(x)$ which is one
realisation of a stochastic process on an infinite domain (i.e.
one for which $x$ can take any value on the real line). The
generalisation of this to a three-dimensional field is entirely
straightforward but the notation is more messy. Now consider the
part of this realisation that lies in the range $-L\leq x < +L$.
We can make a new function of $\delta(x)$ that is periodic by
defining a new function $\delta_L(x)$ such that
\begin{eqnarray}
\delta_L(x) & = & \delta(x)\,\,\,\,\,\, (-L\leq x \leq
+L)\nonumber\\ \delta_L(x+2mL) & = & \delta_L(x) \,\,\,\,\,\,
(m=\pm 1, \pm 2 \pm 3, \ldots)
\end{eqnarray}
The function $\delta_L$ is periodic with period $2L$ so let us
express it as a Fourier series: \begin{equation}
\delta_L(x)=\sum_{n=-\infty}^{+\infty} A_n \exp(ik_nx),
\end{equation}
where $k_n=n\pi/L$ and $A_n$ is given by \begin{equation}
A_n=\frac{1}{2L} \int_{-L}^{+L} \delta_L(x) \exp(-ik_n x).
\end{equation}
Thus, \begin{equation} \delta(x)=\frac{1}{\sqrt{2\pi}}
\sum_{n=-\infty}^{+\infty} G_L(k_n)\exp(ik_n x) \Delta k_n  \label{eq:rep}
\end{equation}
where $G_L(k)$ is defined, for any $k$, via
\begin{equation}
G_L(k)=\frac{1}{\sqrt{2\pi}} \int_{-L}^{+L} \delta_L(x) \exp(-ikx)
dx=\frac{1}{\sqrt{2\pi}} \int_{-L}^{+L} \delta(x) \exp(-ikx) dx
\end{equation}
where $\Delta k_n=k_{n+1}-k_n=\pi/L$. A spectral density function
$P(k)$ can be defined by \begin{equation} P(k)=\lim_{L\rightarrow
\infty} \langle \frac{|G_L(k)|^2}{2L} \rangle
\end{equation}
where the expectation value $\langle \cdot \rangle$ is over the
probability distribution. Note that $|G_L(k_n)|\rightarrow\infty$
as $L\rightarrow\infty$ but $|G_L(k_n)|\Delta(k_n)\rightarrow 0$
as $L\rightarrow\infty$.

Now we define the function $\tilde{\delta}_L(k)$ via
\begin{equation}
\tilde{\delta}_L(k) = \frac{1}{\sqrt{2\pi}} \int_{-\infty}^k
G_L(q)dq, \end{equation} so that
\begin{equation}
\Delta \tilde{\delta}_L(k_n) \equiv
\tilde{\delta}_L(k_{n+1})-\tilde{\delta}_L(k_n) \sim
\frac{1}{\sqrt{2\pi}} G_L(k_n)\Delta k_n
\end{equation}
for small values of $\Delta k_n$, i.e. for large $L$. Hence, in
this limit,
\begin{equation}
\langle | \Delta \tilde{\delta}_L(k_n)|^2 \rangle \sim \langle
|G_L(k_n)|^2 \frac{\Delta k_n}{2\pi} \rangle \Delta k_n = \langle
\frac{|G_L(k_n)|^2}{2L} \rangle \Delta k_n
\end{equation}
and, as $L\rightarrow\infty$ so that $\Delta k_n\rightarrow 0$
\begin{equation}
\langle | \Delta \tilde{\delta}_L(k_n)|^2 \rangle \sim P(k_n)
\Delta k_n \sim \Delta Q(k_n), \label{eq:replim}
\end{equation}
where
\begin{equation}
Q(k)=\int_{-\infty}^{k} P(q)dq
\end{equation}
is the integrated spectrum of $\delta(x)$.

The equation (\ref{eq:rep}) provides a {\em spectral}
representation for $\delta_L(x)$ for all $x$ but which is only
valid for $\delta(x)$ in the range $-L\leq x \leq +L$. To get a
representation that is valid for all $x$ we would like to let
$L\rightarrow \infty$ in (\ref{eq:rep}), but $G_L(k_n)$ does not
converge to a finite value as $L\rightarrow\infty$. However,
$G_L(k_n)\Delta k_n$ is well-behaved so we can write, in the range
$-L\leq x\leq +L$,
\begin{equation}
\delta(x)\equiv \delta_L(x)=\sum_{n=-\infty}^{\infty}
\Delta\tilde{\delta}_L(k_n) \exp(ik_nx).
\end{equation}
Now is the time to let $L\rightarrow\infty$ so that $\Delta k_n
\rightarrow 0$. The sum on the right-hand-side then converges in
the required ``mean-square'' sense to a Stieltjes integral:
\begin{equation}
\delta(x)=\int_{-\infty}^{\infty} d\tilde{\delta}(k)\exp(ikx), \label{eq:square}
\end{equation}
where
\begin{equation}
d\tilde{\delta}(k)=\lim_{L\rightarrow\infty} \Delta
\tilde{\delta}_L(k_n).
\end{equation}
The limit of equation (\ref{eq:replim}) then becomes
\begin{equation}
\langle |d\tilde{\delta}(k)|^2\rangle = dQ(k)=P(k)dk.
\end{equation}
But note that $d\tilde{\delta}(k)$ is {\em not} the Fourier
transform of $\delta(x)$; that role is more properly played by a
term of the form $d\tilde{\delta}(k)/dk$. Following these
considerations we can write
\begin{equation}
\tilde{\delta}(k_1)-\tilde{\delta}(k_2)=\frac{1}{2\pi}
\int_{-\infty}^{+\infty} \left( \frac{e^{-ik_2x}-e^{-ik_1x}}{-ix}
\right) \delta(x) dx.
\label{eq:trippy}
\end{equation}

The expression (\ref{eq:trippy}) has important consequences for
the phases of the Fourier components of the process $\delta(x)$.
First, note that $\langle \delta(x) \rangle=0$. This requires that
$\langle G_L(k) \rangle =0$ which, in turn, means that $\langle
\Delta \tilde{\delta}_L(k)\rangle=0$ and this, in turn, means that
$\langle d\tilde{\delta}_L(k)\rangle=0$ for all $k$. Now, taking
the complex conjugate of equation (\ref{eq:square}) yields
\begin{equation}
\delta^*(x)=\int_{-\infty}^{\infty} d\tilde{\delta}^*(k)\exp(-ikx)
\end{equation}
and replacing $x$ by $x+r$ in equation (\ref{eq:square}) yields
\begin{equation}
\delta(x+r)=\int_{-\infty}^{\infty}
d\tilde{\delta}(k)\exp(ik(x+r)).
\end{equation}
Multiplying these two expressions together and taking the
expectation value leads to
\begin{equation}
\langle \delta^*(x)\delta(x+r)\rangle = \int_{-\infty}^{\infty}
\int_{-\infty}^{\infty} \exp(ix[k'-k])\exp(ik'r) \langle
d\tilde{\delta}^*(k) d\tilde{\delta}(k) \rangle.
\end{equation}
Since $\delta(x)$ is assumed to be statistically homogeneous, i.e.
its probability distribution and all moments are invariant with
respect to translations in $x$, the left-hand-side of this
equation has to be the autocovariance function of the process
$\xi(r)$. The right-hand-side must therefore be a function of $r$
only. The contribution to the double integral must therefore
vanish when $k\neq k'$. In other words,
\begin{equation}
\langle d\tilde{\delta}^*(k) d\tilde{\delta}(k')\rangle=0
\end{equation}
if $k\neq k'$. This tells us immediately that the phases of the
increments of the process, $\arg(d\tilde{\delta}(k))=\phi_k$ must
be random, in the sense that \begin{equation} \langle \exp
i(\phi_k-\phi_{k'}) \rangle = 0  \label{eq:uncorr}
\end{equation}
for $k\neq k'$, since the above argument applies regardless of the
form of $d\tilde{\delta}^*(k)$. The term {\em correlation} is
usually taken to refer to the association expressed by expectation
values like that on the left-hand-side of equation
(\ref{eq:uncorr}), i.e. second-order association. Statistical
homogeneity thus requires that phases of the components of
$\delta(x)$ should be {\em uncorrelated} and is not something to
do with whether the process $\delta(x)$ is Gaussian. Notice also
that the expectation value in equation (42) is an ensemble
average, and this constraint does not apply necessarily to an
average over a finite piece of a single realisation.

Of course the condition (\ref{eq:uncorr}) does not mean that all
phases must be statistically independent; it is weaker than the
requirement that the joint probability distributions of two phases
be separable. Neither does it mean that no statistical quantity
based on the $\phi_k$ recovered from a single realisation can be
found that is sensitive to the non-Gaussianity of $\delta(x)$.
What it does show, however, is that one has to look further than
methods based on second-order properties of the distribution of
phases. This is the challenge we address in the next section.

\section{Quantifying Phase Information}

\subsection{Preamble}
From the examples discussed in Section 3, one  can see that the
signature of non-linear mode coupling is that the phases of the
Fourier modes on large scales couple in certain relationship,
depending on the morphology of the density distribution.  On
smaller scales, however, the phases queue up in such a way that
phase difference, or the discrete phase gradient $D_k$, converges.
This suggests that it might be a good idea to construct a
histogram of the phase gradients for each mode and use it as a
test of phase correlations, interpreted in the wider sense. In a
Gaussian random field, the phases are random, so the phase
gradient will simply be the difference between two
uniformly-distributed random variables and will itself be
uniformly distributed between $-\pi$ and $\pi$. For a non-Gaussian
field, however, this might not be the case. The distribution of
phase differences is therefore one possible way of using phase
information as a diagnostic of non-Gaussianity. Notice that this
histogram is more useful than the histogram of phases themselves,
which would simply be the projection of a return map onto one of
its axes: this would be uniform even in the highly structured
distributions seen in the previous section. Although this
particular approach is suggested by the convergence displayed in
the  simple examples shown above, there could of course be other
forms of phase behaviour that it might also quantify.

Taking a histogram from a set of data produces something
more-or-less equivalent to a probability density distribution. We
can exploit this idea to think of the information content of this
probability density, by defining a quantity analogous to its
entropy. At the start, however, we should stress that this should
by no means be interpreted as a thermodynamic entropy, as it only
relates to the pattern of phases and not to the whole spatial
distribution of material in space.

\subsection{Phase entropy}
We now define a quantity, the {\em phase entropy}, as
\begin{equation}
S=-\sum^{\pi}_{-\pi}\rho_{i}(D_k)  \ln \left[\rho_{i}(D_k)\right]
\delta\phi_{i},
\end{equation}
with
\begin{equation}
\sum \rho_{i}\delta \phi_{i}=1,
\end{equation}
where $\rho_{i}(D_k) \delta\phi_{i}$ is the probability of finding
a value  in the $i$-th bin. This definition takes into account
changes in binning strategy for the construction of the
histograms.  If $\delta \phi_{i}$ were infinitesimal,
\begin{equation}
S=-\int^{\pi}_{-\pi}\rho (D_k) \ln \left[\rho (D_k)\right] d\phi.
\end{equation}
If a data set consisting of $n_{\rm tot}$ values of $D_k$ is
binned into a histogram of $m$ bins, each of which has width
$\delta \phi=2\pi/m$, the probability density for $n$ events in
the $i$-th bin is  $\rho_{i}=n/(n_{\rm tot} \delta \phi_{i})$.

Now another virtue of $D_k$ becomes apparent. The phase of each
Fourier mode changes with the choice of origin, and so does the
converging point of phase gradient. Any statistic we use should
not depend on the choice of origin, or at least this should be
easy to remove from the calculations. Changing the point of origin
by a distance $x$ changes the phase of the $k$-th mode by an
amount $kx$, so the effect on the individual phases depends on
their wavenumber. But the phase gradient suffers a constant offset
\begin{equation}
D_k=\phi_{k+1}+(k+1)x-\phi_{k}+kx=\phi_{k+1}-\phi_{k}+x=D_k+x,
\end{equation}
which can be handled easily. Now take as an example a single
density peak, located at $a$. The phase for the $k$-th mode is
$-ka$, so the phase gradient for all $k$ values of the Fourier
decomposition of the density peak converges to the same value,
\begin{equation}
D_k=\phi_{k+1}-\phi_{k}=-(k+1)a-(-ka)=-a.
\end{equation}

The probability density distribution is the same for all values of
$a$, even with a constant offset caused by different choice of
origin. This approach to phase information weights all phases
equally, and discards power spectral information. As shown in the
previous sections, the way in which the Fourier phases queue up
depends on the relative level of the most dominant density peak to
the rest of the structure:  the higher the dominant peak, the
greater is the effect for large $k$. It is the 'queuing up' of
phases of high-frequency modes that signifies the gravitational
clustering towards the nonlinear regime, quite independently of
the amplitude corresponding to each phase. The quantity $D_k$
encodes information about the ratio of successive Fourier
components:
\begin{equation}
\frac{\tilde\delta_{k+1}}{\tilde\delta_{k}} =
\frac{|\tilde\delta_{k+1}|}{|\tilde\delta_{k}|}
\exp[i(\phi_{k+1}-\phi_k)]=A\exp(iD_k),
\end{equation}
where $A$ is a real number, which is obviously neglected in our
analysis. This is a potential weakness when it comes to applying
these ideas in practice, because a mode with vanishing amplitude
still has a phase which receives equal weight in $D_k$ with those
of much higher amplitude. In real applications, therefore one
would probably want to combine amplitude and phase information in
some way, perhaps by removing modes of very low amplitude from the
analysis. We return to this in Section 7. For the rest of this
paper we shall neglect this problem and look at phase-only
information.

The idea of phase entropy was  suggested by Polygiannakis \&
Moussas (1995). It comes directly from Boltzmann's definition of
entropy in thermodynamics. In an ensemble of $\nu$ identical
replicas, the statistical weight $\Omega_{\nu}$ of its system when
$\nu_{1}$ are in state 1, $\nu_{2}$ in state 2,$\ldots$, and
$\nu_{r}$ in state $r$ is
\begin{equation}
\Omega_{\nu}=\frac{\nu!}{\nu_{1}!\,\nu_{2}!\cdots\nu_{r}!}.
\end{equation}
The entropy of the ensemble is thus \(S_{\nu}\sim\ln
\Omega_{\nu}\). If $\nu$ is sufficiently large, $\nu_{r}$ also
become large, so \(\ln\Omega_{\nu} \sim \nu\ln\nu-\sum_{r}\nu_{r}
\ln \nu_{r}\). Taking $p_{r}=\nu_{r}/\nu$, the probability of
state $r$, we reach \(S \sim -\sum_{r}p_{r}\ln p_{r}\). In the
analogy to the definition of phase entropy, $r$ microstates
correspond to the bins in a histogram, which are between $-\pi$
and $\pi$. Following the work of Shannon (1948a,b) this definition
of entropy $S$ is closely related to the information content $I$
of the distribution: $S(D)=-I(D)$. We do not wish to push the
connection with Boltzmann entropy too far, however, and in any
case the definition of phase entropy is self--contained.
It is not necessary to think about an approximation scheme of
$\Omega_{\nu}$, i.e., requiring $\nu$ to be very large.
Nevertheless, it is still required to have large enough of phases
so as to produce a smooth probability density distribution. In
some cases, where the probability density distributions are highly
skew and therefore some bins in the histogram might have zero
number density, the contribution to phase entropy from these bins
is zero, as \(\lim_{\rho_{r}\rightarrow 0} \rho_{r}\ln
\rho_{r}=0\).

For a Gaussian field, where the number of Fourier modes is
sufficiently large, and the phases of each of these modes are
random, the phase entropy is
\begin{equation}
S=-\sum \rho_{i} \ln \! \rho_{i} \: \delta \phi_{i}=
-\int^{\pi}_{-\pi} \rho(D_k) \ln\left[\rho(D_k)\right] d\phi
=-\int^{\pi}_{-\pi}\frac{1}{2\pi} \ln\left(\frac{1}{2\pi}\right)\;
d\phi= \ln 2\pi.
\end{equation}
It can be proven, by straightforward analogy with the
isoperimetric problem of variational calculus, that phase entropy
has its maximal value when the probability density function
$\rho[D_k]=1/2\pi$ is constant. The Gaussian process, wherein the
distribution of phases (and phase differences) is uniform,
corresponds to a state of maximum entropy, and the value of $S=\ln
2\pi$ is the maximum value that $S$ can take in any situation.

When an initially-Gaussian field evolves under gravitational
instability, the phases 'queue up' in a certain fashion, and the
phase gradient is no longer evenly distributed in the histogram.
What happens then is that the entropy of the phases {\em
decreases} or, in other words, the information content increases.
This does not mean that the process violates the second law of
thermodynamics, because information moves between the amplitudes
and phases (and indeed, between the density and velocity parts of
phase-space) as the system evolves. There is more information in
the whole system than is contained only in the phases, but the
information content of these phases does certainly increase under
the action of gravitational clustering.

Although our application of this idea is largely inspired by
Polygiannakis \& Moussas (1995), who applied it to the analysis of
time-varying quantities in Solar Wind data, they did not take into
account the fact that the straightforward $p\log p$ entropy of the
phase histogram itself is not invariant under translations of the
origin. In our application, we use $S(D)$ rather than $S(\phi)$ to
counter this problem, as above.

\section{Numerical Simulations}

In order to study the behaviour of $S$ in systems with more
complex initial data, we have conducted a series of numerical
N-body experiments. In order to keep these experiments as simple
as possible and to obtain the best possible numerical resolution
with the computer resources available to us, we have used
two-dimensional simulations of $512^2$ particles on a $512^2$ mesh
(like those described in Beacom et al. 1991).

Four different initial pure power law spectra $P(k) \propto k^{n}$
are modelled (with $n=2$, $1$, $0$, and $-1$) for each of which
the initial phases are randomised using a  random number
generator. To see the systematic effects from different power
spectral indices and exclude the influence from phases, we set the
same phases for the same modes for each of the 4 spectral indices.
The boundary of each realisation is periodic. The simulations are
equivalent to 3D Einstein-De Sitter ($\Omega\!=\!1$) universe with
2D perturbations, whose images are cross sections of 3D
perturbations. The spectral index $n$ in 2D is equivalent to $n-1$
in 3D in statistical measure. We also label the evolutionary
stages of the simulations according to $k_{NL}$ where
\begin{equation}
\langle\left(\delta\rho/\rho\right)^{2}\rangle_{k_{NL}}=
b^{2}_{+}(t)\int^{k_{NL}}_{0}P(k)\: d^{2}k =1.
\end{equation}
Here $b_{+}(t)$ is the growing mode in (\ref{eq:linearsol}) of the linear density
contrast, and $P(k)$ is the linear extrapolation of the initial
power spectrum. This  definition of $k_{NL}$ identifies the
corresponding  $2\pi k_{NL}^{-1}$ as the boundary between
linearity and non-linearity. From this criterion and $P(k)
\propto k^{n}$, $k_{NL}^{n+2} \propto b_{+}^{-2}(t) \propto
(1+z)^{2}$. We set the final stage of the simulations the final
stage to have $k_{NL}=2$, and pick 8 stages with values of
$k_{NL}$ varying by a factor of two in each one; the first stage
has $k_{NL}=k_{\rm Nyquist}=256$.

Before proceeding, it is worth illustrating the points we made in
the introduction about the importance of phase information using
these simulations.

To extract a measure of the phase entropy from these
two-dimensional N-body simulations, we simply take the probability
density distribution of all phase gradients in $k_{x}$ direction,
for $S_{x}$, and those in $k_{y}$ direction, for $S_{y}$, then we
take the arithmetic mean, because of the additive nature of the
definition. To express in a clearer way, we can, for example,
write
\begin{equation}
S_{x}=-\sum^{m}_{i=1}\rho(D_{k_x})\ln\left[\rho(D_{k_x})\right]\:
\delta\phi_{x},
\end{equation}
where
\begin{equation}
D_{k_{x}}=\phi(i+1,j)-\phi(i,j) \end{equation} for $1 \leq i,j
\leq 256$, the upper limit being the Nyquist frequency of the
simulations. The number of bins $m$ is chosen according to the
number of phases available, to ensure the probability density
distribution is reasonably smooth. For a realisation of $512^2$
particles on a $512^2$ mesh, there are $256^2$ phases. The phase
gradient $D(k_x)$ for a certain $j$ is the phase gradient for the
corresponding strip in the $x$-direction of the density
distribution in real space. By taking all the phase gradients in
one direction, binning them into a histogram, we are scanning
every strip and examining the morphology of the realisation in
that direction.

\begin{figure}
\epsfig{file=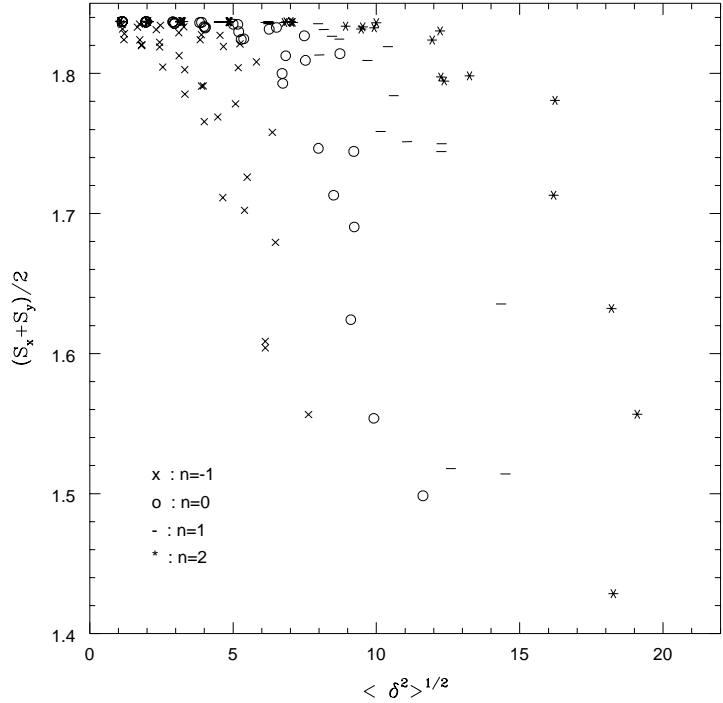,width=10cm} \caption{Phase entropy of
the gravitational evolution from different initial power spectral
indices is shown as a function of the {\em rms} density
fluctuations. For each spectral index, N-body simulation for 5
initial conditions of different random phases are conducted. Phase
entropy, at its maximal value $\ln 2\pi$ at the start, decreases
when the system evolves. The different spectral indices display
different characteristic curves of this decrease in phase
entropy.}
\end{figure}

Phases not only suffer an offset at the choice of origin in real
space, but also depend upon orientation. In reality there should
not be any preferred orientation in which the phase entropy is
calculated. If a large filament lies diagonally across a
realisation, it intercepts with every strip of the realisation at
different positions. The converging points of the phase gradient
for every strip cover the interval $[-\pi, +\pi]$ uniformly, and
each  contributes the same amount of number density to every bin
in the histogram. This results in a large phase entropy. However,
if we rotate the realisation so that the strip lies across the
scanning direction, a lower value of $S$ will be obtained. Here in
Figure~5 all of the realisations are rotated to the orientation
along which the phase entropy is at its minimal value.

We perform 5 sets of different initial phases, so there are 40
realisations for each spectral index. In Figure~5, we plot the
phase entropy against the variance of the density field in the
realisation for all power spectra. (The variance is plotted by
smoothing the density field onto the the grid using a box-counting
algorithm).

It is immediately clear that the phase entropy decreases as the
density variance increases, in other words as the system evolves.
This is in full accord with our intuition described above, and
confirms that $S$ is a potentially useful diagnostic statistic.
Furthermore, the different spectral indices display different
characteristic curves of this decrease in phase entropy. The
trends can be explained qualitatively in terms of the known
morphology of clustering generated by the different sets of
initial conditions. The spectrum with $n=-1$ contains a relatively
large amount of power in long-wavelength modes. The evolution of a
system with this initial spectrum is characterized by the early
generation of long filaments where the caustics form; sheets form
in 3D, but these are not relevant to the 2D simulations used here.
In this case we can always rotate the realisation so that the
longest filament, which has the largest effect on $S$ is parallel
(or perpendicular) to the orientation along which the Fourier
transform is performed. The phase entropy decreases on the
formation of large filaments of size similar to the simulation
box, which happens even when the variance is small. As $n$
decreases so does the amount of large-scale power. In order to
produce the large filaments that result in a decrease of $S$, the
system has to evolve further on small scales. In the most extreme
case of $n=2$, this results in a very high value for the variance
when $S$ starts to display phase structure.

\subsection{Scaling and Self-similarity}

Before further application of phase entropy, we have to address
some important features of the simulations from initial pure power
law spectra. The initial pure power law spectra possess no
characteristic length scales and also evolves in a self--similar
manner, which means that at different stages, the distribution
should possess the same . For a self--similar physical system,
time dependence can be singled out by rescaling ${\bf x}$, the
spatial position, or ${\bf p}$, the momentum. Therefore a
distribution function $f({\bf x},{\bf p},t)$ has the same
statistical measure as the rescaled one, $\lambda^{\alpha}f({\bf
x}/\lambda^{\beta},{\bf p}/\lambda^{\nu},\lambda t)$, as
$t\rightarrow \lambda t$, where $\alpha$, $\beta$, and $\nu$ are
to be decided (Jain \& Bertschinger 1996). Self--similar
gravitational clustering is an idealisation and can not be applied
in detail to a realistic model. It nevertheless provides useful
insights into more realistic situations.

Since the phases of Fourier modes play an important role in
recovering the morphology of a density distribution (Oppenheim \&
Lim 1981), it is interesting to see if there is any self--similar
scaling of the phases in gravitational evolution. As well as the
$512^2$ N-body simulations discussed in the previous section, we
also conduct $2048^2$ simulations with initial pure power law
spectra, $n=2$, $1$, and $0$.  With $k_{NL}=2$ as the final stage
($z=0$), there are 10 stages, denoted stage $a$, $b$ \ldots $j$.
To distinguish  large- and small-scale phases, a realisation of a
$2048^2$ simulation box is divided into $2^2$ ( scale length ${\it
l}_{s}=2048/2=1024$), $4^2$ ($l_{s}=512$), $8^2$ (${\it
l}_{s}=256$), $16^2$ ($l_{s}=128$), etc. boxes (sub-realisations).
On each scale length $l_{s}$, we calculate the phase entropy for
each sub-realisation, and then take average of them over that
scale. By doing so, we are averaging phase entropy for the same
scale length through many sub-realisations. The lower limit of the
scale length of sub-realisations chosen is the one which still has
sufficient large number of phases. Here we choose as the smallest
scale length $l_{s}=64$, by dividing a $2048^2$ realisation into
1024 sub-realisations of a $64^2$ mesh. Each $64^2$ mesh offers
merely $32^2$ phases binned into a histogram of 40 bins.
Therefore, for each $2048^2$ realisation, the phase entropy
appears on 5 different scale lengths, and, from large to small
$l_{s}$, more and more phases from large scales  are
excluded from the calculation. It is expected that phase entropy
decreases through large to small scales, as structure on small
scales goes nonlinear while that on large scales is still in the
linear regime. To simplify the calculations and for completeness,
the sub-realisations are not rotated to minimise the  phase
entropy in this case. Taking $l_{s}=512$ as an example, we divide
a realisation of a $2048^2$ simulation box into $(2048/512)^2=16$
sub-realisations, each of which is a $512^2$ mesh, and calculate
the phase entropy for each sub-realisation, before taking average
of the phase entropy of these 16 sub-realisations. Structure
scales larger than $512$, $1/4$ of the side of the whole
simulation box, are left out, what is included in the calculation
being the structure up to the scale length $l_{s}$.

\begin{figure}
\epsfig{file=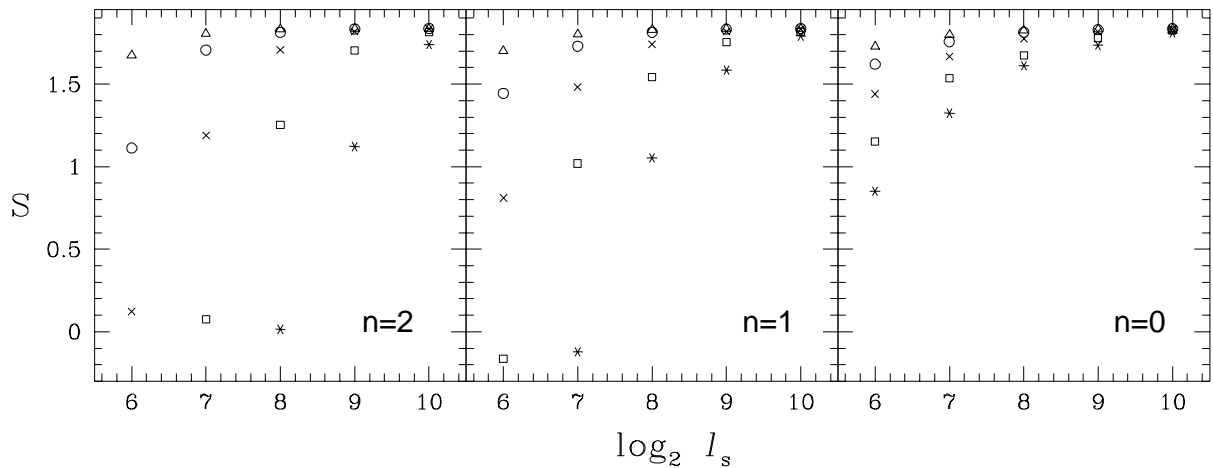,width=17cm} \caption{The phase
entropy is  shown for five length scales $l_{s}=64$, 128, 256, 512
and 1024. $\triangle$ denotes stage $f$, $\circ$ denotes stage
$g$, $\times$ stage $h$, $\Box$ stage $i$, and  $\ast$ stage $j$.
Clearly, phase entropy decreases towards small length-scale as
expected. For the same stage of $n>0$, the transition in scale
from linearity to nonlinearity can be seen from the large decrease
in phase entropy by decreasing $l_{s}$, e.g., for stage $g$ from
$l_{s}=128$ to 64, stage $h$ from $l_{s}=256$ to 128, stage $i$
from $l_{s}=512$ to 256, and stage $j$ from $l_{s}=1024$ to 512.
Also phase entropy seems an indicator to nonlinearity of
gravitational clustering. The amount of decrease in phase entropy
is roughly the same for the transition from linearity to
nonlinearity for every stage, which implies self-similar nature of
gravitational clustering for pure power law spectra.}
\end{figure}

In Figure~6, the phase entropy is shown on 5 scale lengths for
stage $f$ to $j$. The phase entropy decreases from large to small
$l_{s}$ for all three spectral indices. For stage $g$ of spectral
index $n=2$, the phase entropy decreases drastically from scale
length $l_{s}=128$, to $l_{s}=64$, which can be viewed as follows.
The stages of simulations are chosen to indicate the boundary
between linearity and nonlinearity and that there is factor of two
between any two adjacent stages. For stage $g$ ($k_{NL}=16$), the
boundary is on a box size of $(2048/16)^2=128^2$, meaning that
structure within a $128^2$ box placed anywhere in the simulation
box has gone nonlinear in a statistical sense. Thus the transition
from linearity to nonlinearity across the boundary is shown
clearly via the large decrease of the phase entropy from length
scale $l_{s}=128$ to $l_{s}=64$. The same situation applies to
stage $h$ from $l_{s}=256$ to $l_{s}=128$, where $l_{s}=256$
corresponds to the boundary size of nonlinearity; so does stage
$i$ from $l_{s}=512$ to $256$, and stage $j$ from ${\it
l}_{s}=1024$ to $512$. What is more interesting in Figure~6 is
that phase entropy seems to stay at the same level for the same
degree of nonlinearity in the plot. Stage $g$ at scale length
$l_{s}=64$ possesses the same level of phase entropy as stage $h$
at $l_{s}=128$, so does stage $i$ at $l_{s}=256$, and stage $j$ at
$l_{s}=512$. If $k_{NL}$ is used to denote the stages,
\(S(k_{NL},l_{s}) = S(2,512) \sim S(4,256) \sim S(8,128) \sim
S(16,64)\). This result is not surprising at all. As $k_{NL}$ are
chosen twofold for two adjacent stages in the self--similar
evolution, the distribution of stage $m$ looks the same
statistically as stage $m+1$ when blown up by a factor of two. For
any scale length $l_{s}$ of stage $m$, it should retain the same
level of phase entropy as the next stage $m+1$ at double scale
length, $2l_{s}$. Therefore, the phase entropy is indeed an
indicator of the level of nonlinearity (or linearity), and
demonstrates the roughly self--similar nature of the evolution,
which can be expressed as
\begin{equation}
S(k_{NL},l_{s})=S(2k_{NL},l_{s}/2),
\end{equation}
as $k_{NL} \rightarrow 2k_{NL}$, or equivalently, $(1+z)
\rightarrow 2^{\frac{n+2}{2}}(1+z)$. All the stages can be scaled
by phase entropy into one single characteristic curve,
$S(k_{NL}l_{s})$.

This simple picture is complicated by a number of factors which
produce a discernible slope in Figure 6, which displays a
perceptible tilt toward small  $l_{s}$. This disagreement between
points of the same level of nonlinearity may be due to the fact
that $S$ does not scale in a self-similar fashion. But before this
can be concluded we have to contend with several complicating
factors. First, for large $l_{s}$, there are fewer
sub-realisations to be averaged, and the sub-realisations in this
experiment are not rotated  to obtain the value of the phase
entropy. This is expected to artificially increase the phase
entropy on large-scales compared to small. This is probably more
pronounced for small $n$ where filaments dominate. For small
$l_{s}$ the averaging of more sub-realisations erases this effect.
Using larger bins to compensate for the smaller number of phases
available also lowers the phase entropy. We therefore see points
at the same level of phase entropy tilt towards small length
scales. Another numerical effect comes in the when both  $k_{NL}$
and $l_{s}$ are small. In forming clusters, galaxies aggregate at
the expense of forming voids. When the evolution goes into the
highly nonlinear regime, voids dominate the realisation and many
of the sub-realisations become empty. This produces an artificial
contribution to the entropy determined partially by the bin size.

But while the exact form of scaling remains to be determined by
eliminating these possible systematics and studying the scaling of
$S$ for a wider range of spectral indices, this study at least
shows that phase entropy has qualitatively the right properties
for a diagnostic of gravitational evolution.

\section{Discussion and Conclusions}

The aim of this paper was to explore the behaviour of phases of
Fourier components of cosmological density fluctuations in order
to understand how phase information relates to the development of
clustering pattern from random-phase (Gaussian) initial
conditions.

We have studied some simple cases of the evolution of Fourier
phases under the action of gravitational instability and used
these cases to motivate the definition of a measure of the
information contained in these phases. The statistic we propose,
$S$, which behaves as a kind of information entropy, shows
interesting time-evolution and appears to undergo a form of
scaling when the initial conditions on which gravity acts are
self-similar.

Although this statistic looks like a promising candidate as a
diagnostic for phase coupling, there are many problems to be
overcome before it can be applied to real data. For example, it
remains to be seen how $S$ can be corrected for shot-noise
(discreteness) errors and also how to correct for the imposition
of a selection function. Redshift-space effects also need to be
taken into account. These are inevitably more difficult to handle
for a higher-order statistic than with the correlation function or
the power-spectrum. There is also the problem, alluded to in
Section 5, that even Fourier modes with negligible amplitudes have
a phase, and our statistic does not take account at all of
amplitude information. In a practical application one would
probably therefore prefer to combine amplitude information with
phase information. This is what is done to some extent with the
bispectrum (Matarrese et al. 1997) and the second spectrum
(Stirling \& Peacock 1996). But note that the bispectrum, for
example, only contains information about third-order moments of
the Fourier components. It does not therefore include all the
information that resides in the distribution of phases. Knowledge
of all higher-order polyspectra would be necessary to define the
total information content of the Fourier phases. For this reason
it is certainly worth investigating statistics based explicitly on
the phases, rather than indirectly on them as is done with the
bispectrum or any other piecemeal approach higher-order moments.

In future we hope to test the usefulness of phase statistics
compared with more standard methods. In any case we stress the
point that phase information is completely complementary to
amplitude information: no information about localised geometrical
structures is contained in the amplitudes themselves, so phase
information however it is encoded is vital to a full analysis of
clustering pattern.

As we discussed in the introduction, phase information {\em per
se} has not been extensively used in galaxy clustering analyses so
far. We hope to develop tools that use this information to provide
sensitive tests of cosmological models. In the context of galaxy
clustering, what we need to understand using further N-body
experiments is how the growth of phase information relates to the
growth of the power spectrum. It is known that gravitational
instability acts on initially Gaussian perturbations to produce
fluctuations that are non-Gaussian but which are characterised by
higher-order moments, such as the skewness $\langle \delta^3
\rangle$ that couple to the variance $\langle \delta^2 \rangle$ in
a well-defined way (e.g. Coles \& Frenk 1991; Bernardeau 1992).
One of the possibilities we shall explore is how the growth of
phase information itself couples to the growth of the Fourier
amplitudes. If this coupling can be quantified, then one can hope
to use it as a diagnostic of clustering. For example, many
clustering models appeal to the presence of a bias to enhance the
clustering pattern over that generated by gravity (e.g. Coles
1993). A linear bias $b$ affects the Fourier amplitudes but not
their phases. One can therefore hope to distinguish a biased power
spectrum from one generated solely by gravitational instability:
for the same amplitude, the former should have a higher phase
entropy (lower phase information) than the latter.  By quantifying
the phase information on very large scales we also hope to test
initial non-Gaussianity, if the non-Gaussianity developed from
initially Gaussian fluctuations can be understood. We shall return
to these and related questions in future work, but the scaling
described in Section 6 is grounds for optimism that this can be
achieved.

We also take this opportunity to point out that the phase entropy
can in principle be applied not just to density fluctuations, but
also as a diagnostic of non-Gaussianity in microwave background
temperature fluctuations. With tentative claims already being made
for the detection of non-Gaussian effects in the COBE data in an
analysis based on bispectra (Ferreira, Magueijo \& Gorski 1998),
though not without opposition (Bromley \& Tegmark 1999),  the
pre-Planck development of sensitive yet robust statistical
indicators is a matter of urgent priority. Phases can be defined
for a spherical harmonic expansion in a similar way to that of the
Fourier modes and may provide more information about the form of
non-Gaussian behaviour present in existing CMB maps as well as
offering more powerful and sensitive descriptors of future data
sets.

\end{document}